# Smart materials and structures for energy harvesters

**Tian Liu[1], Sanwei Liu[1], Xin Xie[1], Chenye Yang[2], Zhengyu Yang[3], and Xianglin Zhai[4]***

[1] *Department of Mechanical and Industrial Engineering, Northeastern University, Boston, MA 02115*
[2] *Department of Electrical Engineering and Computer Science, Massachusetts Institute of Technology, Cambridge, MA 02194*
[3] *Department of Electrical and Computer Engineering, Northeastern University, Boston, MA 02115*
[4] *Department of Chemistry, Louisiana State University, Baton Rouge, LA 70803*

* alex.zhai@poochonscientific.com

## Abstract

Vibrational energy harvesters capture mechanical energy from ambient vibrations and convert the mechanical energy into electrical energy to power wireless electronic systems. Challenges exist in the process of capturing mechanical energy from ambient vibrations. For example, resonant harvesters may be used to improve power output near their resonance, but their narrow bandwidth makes them less suitable for applications with varying vibrational frequencies. Higher operating frequencies can increase harvesters' power output, but many vibrational sources are characterized by lower frequencies, such as human motions. This paper provides a thorough review of state-of-the-art energy harvesters based on various energy sources such as solar, thermal, electromagnetic and mechanical energy, as well as smart materials including piezoelectric materials and carbon nanotubes. The paper will then focus on vibrational energy harvesters to review harvesters using typical transduction mechanisms and various techniques to address the challenges in capturing mechanical energy and delivering it to the transducers.

**Ambient energy sources**

Energy harvesters can extract energy from ambient sources, convert it to electrical energy, and use it to power electronic devices wirelessly. Energy harvesters based on solar energy [Tan 2011, Abdin 2013], thermal energy [Harb 2011], electromagnetic energy [Pinuela 2013], and mechanical energy [Saadon 2011] have been reported recently. Among many review papers, including [Steingart 2009, Harb 2011, Watral 2013], Harb et al. describes many harvesters that use the energy sources mentioned above and concludes that vibrations are the most available energy source and provide the highest output powers [Harb 2011].

**Solar energy harvesting**

The sun supplies energy to the planet, and solar harvesting captures some of that sunlight directly and converts it to electricity. Solar technologies including solar photovoltaics (PV) and concentrating solar power (CSP) are widely used to convert sunlight into electrical energy. A recent review of solar technologies and applications can be found in [Devabhaktuni 2013].

One example is found in [Roundy 2003], in which a self-contained 1.9 GHz radio frequency transmit beacon is reported. The device is powered by a cadmium telluride (CdTe) solar cell that is attached on the back side of a PCB and charges a capacitor on the front. The solar powered beacon achieves 100% duty cycle under high light conditions.

Solar cells based on the photoelectric effect are suitable for applications relevant to energy harvesters such as wireless sensor networks and portable electronic devices. However, the availability of solar energy is unpredictable due to its high dependence on weather conditions. Moreover, devices that are required to be installed under clothes or inside the human body can hardly be charged by solar energy.

**Thermal energy harvesting**

Thermal gradients arise as a result of many processes, including the burning of fuel and geothermal processes. Thermoelectric energy harvesters extract electrical energy from temperature gradients. The energy extraction is enabled by the Seebeck effect, in which the temperature difference across a conductor or semiconductor results in a net flow of electrons from the high-temperature side to the low-temperature side so that an electrical current (and voltage) are generated.

One example of thermoelectric harvesting may be found in [Kishi 1999]. Kishi et al. developed a thermoelectric device that can power a wrist watch. With a temperature difference of 2°C between the high-temperature substrate and the low-temperature substrate, the thermoelectric device can produce an power output of 5.6 mW with a load resistance of 1 k$\Omega$. A wrist watch can be driven by connecting 16 such thermoelectric devices in series at room temperature.

However, the performance of such thermoelectric energy harvesters is limited by the available temperature difference of the application and the Carnot efficiency. The requirement of high temperature difference for high power output conflicts with the low temperature difference available in many applications, such as harvesting energy from the human body.

**RF energy harvesting**

In the presence of electronics or radiated power sources such as TV/radio transmitters and mobile base stations, there are typically electromagnetic fields. Another group of energy harvesters can generate electrical power by capturing energy from the electromagnetic field in the free space. The high-frequency electromagnetic energy can be converted into electrical current by using a dipole antenna to capture the electromagnetic waves and a diode connected across the dipole to rectify the AC current into DC current. Aparicio et al. and Lu et al. recently present a good introduction and review of such energy harvesters [Lu 2015, Aparicio 2016].

Although electromagnetic energy sources usually have low power densities (0.01-0.1 $\mu W/cm^2$) as compared with other sources [Lu 2015], the power in free space can be useful for remotely distributed sensors. However, in order to capture enough electromagnetic energy from waves in free space, independent antennas and matching circuits need to be specifically designed, which is challenging for electronics requiring small device volumes.

**Vibrational energy harvesters**

This thesis concentrates on vibrational energy harvesters. Vibrational energy harvesters can be made by using resonant systems or non-resonant systems. An example of a non-resonant harvester is found in [Bowers 2009]. Bowers et al. developed a non-resonant system in which a magnetic ball moves inside a human-carried, coil-wrapped sphere, thereby generating power. The non-resonant design avoids the typically higher resonant frequencies of smaller systems. Resonant energy harvesters are numerous and typically consist of a mass-spring-damper oscillator integrated with a transducer that converts mechanical energy into electrical energy.

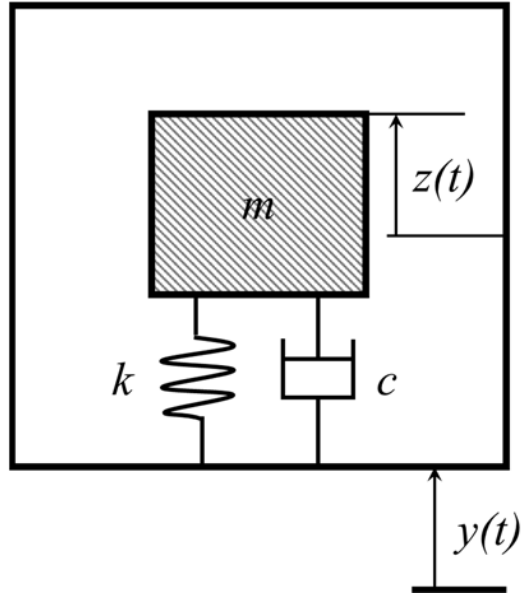

**Figure 2.1.** Schematic diagram of a single degree of freedom lumped parameter model for vibrational energy harvesters.

The simplest resonant harvester systems can be modeled as a single degree of freedom lumped spring-mass-damper system [Williams 1996], as shown in figure 2.1. The model consists of a proof mass *m* connected to the vibrating platform by a spring *k*, and a damper *c* representing the energy dissipation in mechanical form and electrical form. The governing equation of motion for such one-dimensional spring-mass system can be written as

$$m\ddot{z}(t) + c\dot{z}(t) + kz(t) = -m\ddot{y}(t) \tag{2.1}$$

where *y(t)* is the displacement of the platform and *z(t)* is the relative motion of the proof mass with respect to the platform.

In equation (2.1), the converted energy is modeled as the energy dissipated through the electrical damping. In a conventional spring-mass-damper system such as the one described here, the dissipation is proportional to the velocity. This model is accurate for harvesters using electromagnetic transducers; for example, the velocity of a magnet can drive changes in magnetic flux through a coil, which then induce voltages. For some other transduction mechanisms, such as piezoelectric transducers, the energy conversion is proportional to displacement rather than velocity. However, although the spring-mass-damper model may not accurately describe the energy conversion for all transducers, results from this model are often used for analysis of harvesting performance.

For a sinusoidal excitation $y = Y_0 \sin(\omega t)$, the generated power from the vibrational energy harvester can be derived as

$$P = \frac{m \zeta_e \omega_n \omega^2 \left(\dfrac{\omega}{\omega_n}\right)^3 Y_0^2}{\left(2\zeta_T \dfrac{\omega}{\omega_n}\right) + \left(1 - \dfrac{\omega^2}{\omega_n^2}\right)^2} \qquad (2.2)$$

where $Y_0$ is the amplitude of the acceleration on the proof mass, $\zeta_T$ is the total damping coefficient including mechanical damping and electrical damping, and $\omega_n$ is the resonant frequency of the system. The power output reaches its maximum value when the excitation frequency equals the resonance frequency. Thus the power output at the harvester's resonance frequency can be derived as

$$P = \frac{m Y_0^2 \omega_n^3}{4(\zeta_m + \zeta_e)}, \qquad (2.3)$$

where $\zeta_m$ is the mechanical damping coefficient. Equation (2.3) shows that the power output at resonance frequency is proportional to the amplitude of excitation squared.

Mitcheson et al. derived the maximum converted energy of a resonant harvester by multiplying the total distance $4Z_l$ traveled during an oscillating cycle by the force $m\omega^2 Y_0$ applied on the proof mass to obtain a maximum converted energy of $4Z_l m\omega^2 Y_0$ [Mitcheson 2008]. The maximum power can then be derived by dividing the maximum energy by the excitation period $\frac{2\pi}{\omega}$ as

$$P_{max} = \frac{2}{\pi} Y_0 Z_l \omega^3 m. \tag{2.4}$$

Although the equation above overestimates the maximum converted power (because the acceleration involved is sinusoidal instead of constant at the maximum value), the expression clearly indicates that the power generation is limited by the oscillator's maximum displacement.

*Transduction mechanisms of vibrational energy harvesters*

*i Electromagnetic energy harvesters*

A vibration-based electromagnetic energy harvester produces power via the relative movement between a permanent magnet and a coil. Such a system can be effectively realized by a resonating spring or cantilever beam, in which either the permanent magnet or the coil can be chosen to be mounted on the cantilever while the other stays fixed. The voltage output of the coil is determined by Faraday's Law as

$$\varepsilon = -\frac{d\Phi_B}{dt}, \qquad (2.5)$$

where $\varepsilon$ is the electromotive force (EMF) and $\Phi_B$ is the magnetic flux. The key parameters affecting the generated power are the strength of the magnetic field, the relative velocity between the magnet and coil, and the number of turns of the coil.

Amirtharajah et al. reported a self-powered signal processor based on electromagnetic energy harvesting [Amirtharajah 1998]. In their device, a coil is mounted on the proof mass of a mass-spring oscillator with a permanent magnet fixed below. The transducer generates electrical current in the coil when the proof mass vibrates above the permanent magnet. The harvester was estimated to have an average power output of 400 µW under a stochastic excitation simulating human walking.

More recently Ylli et al. developed an electromagnetic energy harvester that is excited by the shock produced by a heel strike [Ylli 2015]. In their device, a metal arm is mounted on a pivot; a magnet is attached at the free end of the arm. The arm's free end is suspended between two additional magnets, one above and one below the free end of the arm. The electromagnetic transducer (magnetic circuit) is assembled near the free end of the arm. The device achieves a maximum average power output of 4.13 mW at a walking speed around 5 km/h with a device volume of 48 cm$^3$. Other examples of harvesters using electromagnetic transducers can be found in [Shearwood 1997, El-hami 2001, Spreemann 2012, Salauddin 2016].

*ii Electrostatic energy harvesters*

An electrostatic energy harvester usually consists of a fixed electrode plate and a movable electrode plate with a separation between the two. Charges are stored between the two plates by connecting the harvester to a voltage source before operation. The voltage across the capacitor is

$$V = \frac{Q}{C}, \tag{2.6}$$

in which $Q$ represents the charge and $C$ represents the value of capacitance. The capacitance $C$ can be expressed as

$$C = \frac{\varepsilon_0 A}{D}, \tag{2.7}$$

in which $\varepsilon_0$ is the dielectric constant of free space, $A$ is the overlap footprint area of the two plates forming the capacitor, and $D$ is the vertical distance between the two plates.

The generated power comes from the work done against the electrostatic force when relative movement between the two electrode plates changes the capacitance. The electrostatic force depends on the capacitor's geometry and the amount of charge stored. Mitcheson et al. concluded that at the micro-scale, electrostatic transducers become more suitable than electromagnetic transducers as they are more compatible with MEMS fabrication and produce greater power densities [Mitcheson 2008].

An example of an electrostatic energy harvester is found in [Meninger 2001]. The harvester consists of a proof mass with comb fingers on both sides, a pair of folded springs connecting the mass to the anchors, and two stationary combs interdigitating with the mass' combs. When the mass oscillates, the capacitance between the two pairs of combs changes, which generates electrical current. The device was predicted to achieve a power output of 8 µW. Such a system's power output can be improved by increasing the capacitance change during mass oscillation, for example by using longer comb fingers or by adding more fingers.

*iii Piezoelectric energy harvesters*

Piezoelectric materials can be used as motor and actuators [Xie 2014]. Energy harvesters using piezoelectric transducers have drawn considerable attention recently. Thorough reviews of piezoelectric energy harvesters are given in [Kim 2007, Saadon 2011]. Piezoelectric transducers are attractive for energy harvesters as the transduction mechanism is simple. A piezoelectric material can generate an electric field when an external mechanical force is applied.

Roundy et al. presented a particularly well-known early example of such a piezoelectric energy harvester using a piezoelectric cantilever [Roundy 2004]. In their device, the cantilever is made of two piezoelectric layers with a metal shim in between. A metal block is simply attached at the free end of the cantilever as a proof mass. When accelerations couple into the proof mass, the piezoelectric beam bends, inducing bending stresses and voltages in the piezoelectric material. An experimental prototype with an overall volume of 1 cm$^3$ achieves a power output of 375 µW under a driving frequency of 120 Hz and an acceleration of 0.26 g.

*Energy capture by vibrational energy harvesters*

In vibrational energy harvesters, mechanical energy is converted into electrical energy in two steps. First, mechanical energy is captured from ambient vibration and delivered to the transducer. Second, the transducer converts the delivered energy from the mechanical domain into the electrical domain. Regardless of the chosen type of transducer, challenges often exist in the first step, which limits the frequency range and power output of such energy harvesters.

As discussed in the context of equation (2.2), resonant harvesters only have large displacements when the excitation frequency is at their resonance frequencies. Resonant systems usually have narrow bandwidths of harmonic oscillation, so that such harvesters can only have improved power output in a small frequency range. This makes such harvesters challenging for applications in which the ambient frequency varies over time.

Another challenge is that the maximum power of a resonant harvester is limited by the maximum displacement of its proof mass and increases with higher excitation frequency as shown in equation (2.3). For a resonant system with a fixed resonance frequency, increasing the excitation frequency beyond the resonance will quickly reduce the displacement of the proof mass. On the other hand, increasing the maximum possible displacement of the structure will inevitably decrease the resonance frequency of the system. Hence it is important to find a solution in which resonant harvesters can be driven at a low resonance frequency but are capable of generating power at a high frequency.

*i Frequency up-conversion*

For resonant systems, the technique called frequency up-conversion has been extensively studied recently [Kulah 2008, Jung 2010, Pozzi 2011, Gu 2011, Tang 2011, Pillatsch 2012, Zhu 2015]. In frequency up-conversion, the system couples into low frequency ambient excitations and delivers some of the captured energy to the parts of the transduction mechanism, which has a higher characteristic frequency of operation. The system usually consists of a low-resonance driving element that couples to the low ambient driving frequency and a generating element that has high resonance frequency for increased power output. Low-frequency ambient motions can be converted to higher frequency motions of the generating element via the interaction between the two elements. Frequency up-conversion can decrease resonant systems' high resonance frequency to permit large internal displacements of the driving element while retaining efficient power output at the transducer's high resonance frequencies.

Frequency up-conversion can be accomplished by physical contact (usually impact) of a driving element with the generating element [Pozzi 2011, Gu 2011, Jung 2010, Liu 2014, Liu 2015] or through non-contact interactions such as magnetic force [Kulah 2008, Tang 2011, Zhu 2015]. For both types of interactions, the operation mode of frequency up-conversion can be categorized based on the type of interaction dynamics between the driving element and the generating element. In some cases [Gu 2011], the driving element and generating element bounce off each other. In other cases [Pozzi 2011], the driving element and the generating element pass each other, so that the generating element is plucked. Such a harvester's performance largely depends on the dynamics of frequency up-conversion.

Gu et al. describe the modeling and experimental demonstration of a harvester using coupled-motion frequency up-conversion [Gu 2011]. The system has a low-frequency driving beam that impinges on a higher-frequency piezoelectric generating beam. After impact, the driving beam bounces off the generating beam. During the bounce, the beams move together in something called coupled motion. After the bounce, the beams move separately. The repeated bouncing leads to recurring cycles of coupled and individual motion of the two beams, similar to a person bouncing on a trampoline. In coupled-motion frequency up-conversion, power is generated primarily during the two beams' coupled motion and exceeds that produced by a comparably-sized, conventional single-beam harvester.

An example of plucked frequency up-conversion harvesting is demonstrated in [Pozzi 2011]. A set of PZT bimorph beams are mounted on a rotor. When the rotor rotates, the PZT beams move relative to plectra that are mounted on a surrounding stator. As the harvester operates, each PZT bimorph is repeatedly plucked by the plectra mounted on the stator. When a plectrum moves past a PZT generating beam, it first deflects it, then moves past it and releases it to ring down. This architecture enables the mechanical energy that drives the stator's low frequency rotational motion to be extracted by the generating beams at their higher resonance frequency. Pillatsch et al. reported another example of plucked harvesting in [Pillatsch 2012]. In that device, a metal cylinder rolls inside a track above an array of piezoelectric beams with magnets at their tips. Plucking is achieved by the magnetic force between the metal cylinder and the tip magnets of the generating beams.

In frequency up-conversion, the use of a low-frequency driving element reduces the system's resonance for better coupling into human-induced motions. However, the large displacement of the driving element significantly increases the operating volume of the harvesting system. This issue can be improved by separating the proof mass from the transducer to form a non-resonant system.

Minh et al. demonstrated an impact-based piezoelectric energy harvester [Minh 2015]. Their device consists of a piezoelectric generating beam above a metal ball guided in a cylindrical cavity. The motion of the metal ball is limited by the distance between the generating beam and the bottom of the cavity. The power output of their experimental harvesters increases in a wide frequency range from 20 Hz to 150 Hz under an acceleration of 4g.

*ii Resonance broadening*

As a resonant system usually has a single harmonic with a narrow bandwidth [Qian 2013, Cassella 2016], broadening a resonant harvester's frequency range for improved power output is highly desired. Resonant harvesters' bandwidth can be extended by using nonlinear systems. Nonlinear systems can be achieved by adding nonlinear elements such as nonlinear springs or motion stops into the resonating system. Thorough reviews of such energy harvesters using nonlinear systems are given in [Zhu 2010, Harne 2013].

One approach to adding nonlinear elements is to make the spring k in the spring-mass system nonlinear. This can be realized by replacing the traditional spring with a nonlinear Duffing spring. For Duffing springs, the spring force behaves as $F = k_1 x + k_3 x^3$, in which x represents the displacement of the proof mass. When the displacement is small, the linear part $k_1 x$ dominates the spring force. When the displacement increases, the nonlinear term $k_3 x^3$ begins to dominate the spring force, causing the resonance to shift towards the driving frequency.

An example of resonant harvesters using nonlinear springs can be found in [Mann 2009]. The harvester has a center magnet as the proof mass suspended by two side magnets fixed on the top and bottom of the support, both repelling the center magnet. The magnetic restoring forces between the side magnets and the center magnet perform as the nonlinear spring. More recently, Salauddin et al. presented an electromagnetic energy harvester using a magnetic spring as well [Salauddin 2016]. The system's resonance shifts with the driving frequency, increasing the frequency range of improved power.

Another approach to broaden resonant harvesters' frequency range is to build a nonlinear system by using bistable systems. A bistable system has two statically stable equilibrium states. The inertial mass either oscillates around one stable state or snaps back and forth between the two stable states depending on the excitation amplitude [Harne 2013].

Erturk et al. exploited a bistable energy harvester configuration based on a piezomagnetoelastic structure [Erturk 2009]. Their device consists of a ferromagnetic cantilever with two piezoelectric pads fixed near the mounting position. Two magnets with the same direction of polarization are installed on the platform with equal offset from the center, attracting the cantilever. The system is designed to have two stable equilibrium positions. Modeling and experiment both show that the device performs periodic oscillations with large amplitude over a wide frequency range. The harvester achieves a 200% larger open circuit voltage as compared with a harvester without bistability.

Besides nonlinear springs and bistable system, nonlinear systems can also be achieved by using mechanical motion stops, also known as end stops. Le et al. extended the bandwidth of an electrostatic energy harvester by adding end stops to limit the proof mass motion [Le 2012]. In their device, protrusions on the four fixed anchors serve as the end stops for the proof mass. When the amplitude of excitation increases or the driving frequency gets close to the resonance frequency, the proof mass impacts the end stops so that the device behaves nonlinearly. With motion stops, such a harvesting system usually performs a voltage or power saturation phenomenon (power plateau) around the resonance frequency, instead of the voltage or power peak observed for a typical linear system without motion stops. Although the peak power is decreased due to the power plateau, the frequency range is significantly increased.

*iii Resonance tracking*

Resonant harvesters' resonance frequencies can also be tuned to match a time-varying driving frequency. The narrow bandwidths of resonant energy harvesters can be extended by tuning key elements of the mass-spring system, namely the stiffness of the spring, to match the system resonance to the varying driving frequency. Many examples are given in [Zhu 2010, Harne 2013] and the references therein. Among these techniques, a harvester's bandwidth can be actively tuned by applying external tuning input, such as an axial preload, to adjust the resonance frequency to match the driving frequency, by either manual means or a sensor-controlled tuning system.

Eichhorn et al. developed a piezoelectric energy harvester in which the resonance frequency is tuned by applying mechanical stress onto the structure [Eichhorn 2009]. The preload is controlled by manually adjusting the displacement of the spring, and the resulting axial force is delivered by the two arms connecting the free end of the harvesting beam. The resonance frequency of the resulting device can be tuned in the range of 292 Hz-380 Hz with compressive loading and 440 Hz-460 Hz with tensile loading.

The energy required for active tuning can also be from the energy generated by the harvester itself. For example, Eichhorn et al. reported another piezoelectric energy harvester that can self-tune its resonance frequency [Eichhorn 2011]. The structure consists of a piezoelectric generator and a piezoelectric actuator applying axial load onto the generator. The device also has a control unit that decides the amount of harvested power transferred to the actuator based on the analysis of the ambient vibration frequency. Although the minimal power required is quite low at 10% of the generated power, some power will always be required to drive this type of active tuning. Other examples of actively-tuned energy harvesters include [Leland 2006, Peter 2009].

Passive tuning, or self-tuning, does not consume extra energy. Passively tuned systems leverage a system's inherent nature to adapt its resonance frequency to the ambient frequency without external intervention. Gu et al. developed self-tuning energy harvesters for rotating applications [Gu 2010, Gu 2011]. A composite beam of piezoelectric material and plastic is mounted along the radial direction of a rotating platform. The cantilever's stiffness is tuned by the tensile stress caused by the varying centrifugal force. This self-tuning harvester achieves a bandwidth of 8.2 Hz with a center resonance frequency of 13.2 Hz.

Another example of a passively-tuned harvester is found in [Miller 2013], in which a self-tuning phenomenon was discovered for a resonant harvester. In their device, a slider, as the proof mass, can move freely on a double-clamped piezoelectric beam. When the driving frequency changes, the proof mass can slide along the beam adapting the resonance frequency to the driving frequency. All of the systems demonstrated in [Miller 2013] achieve wide bandwidth between 6 Hz and 40 Hz. Other examples of passively-tuned energy harvesters can be found in [Marzencki 2009, Wang 2012, Liu 2012].